\documentclass[american,aps,pra,reprint,superscriptaddress,twocolumn]{revtex4-1}
\usepackage[unicode=true,pdfusetitle, bookmarks=true,bookmarksnumbered=false,bookmarksopen=false, breaklinks=false,pdfborder={0 0 0},backref=false,colorlinks=false] {hyperref}
\hypersetup{ colorlinks,linkcolor=myurlcolor,citecolor=myurlcolor,urlcolor=myurlcolor}
\usepackage{braket,colortbl,amsthm,amsmath,amssymb,cleveref,txfonts}
\definecolor{myurlcolor}{rgb}{0,0,0.7}
\usepackage{graphics,graphicx}
\usepackage{color}

\theoremstyle{plain}

\def\bea{\begin{eqnarray}}
\def\eea{\end{eqnarray}}
\def\ba{\begin{array}}
\def\ea{\end{array}}

\def\ket{\rangle}
\def\bra{\langle}
\def\beq{\begin{equation}}
\def\eeq{\end{equation}}

\begin{document}

\title{ Detecting non-Markovianity via uncertainty relations}
\author{Ananda G. Maity}
\email{anandamaity289@gmail.com}
\affiliation{S. N. Bose National Centre for Basic Sciences, Block JD, Sector III, Salt Lake, Kolkata 700 106, India}
\author{Samyadeb Bhattacharya} 
\email{sbh.phys@gmail.com}
\affiliation{S. N. Bose National Centre for Basic Sciences, Block JD, Sector III, Salt Lake, Kolkata 700 106, India}
\affiliation{Centre for Quantum Optical Technologies IRAU, Centre of New Technologies, University of Warsaw, Banacha 2c, 02-097 Warsaw, Poland }
\author{A. S. Majumdar}
\email{archan@bose.res.in}
\affiliation{S. N. Bose National Centre for Basic Sciences, Block JD, Sector III, Salt Lake, Kolkata 700 106, India}

\begin{abstract}
We present a formalism for detection of non-Markovianity through uncertainty relations. We show that when there is an information back-flow to the system from its environment through CP-divisibility breaking, the Choi-states corresponding to the reduced system evolution contain at least one negative eigenvalue. The consequent break down of uncertainty relations for such states can be used to witness non-Markovian dynamics. We present some relevant examples of the phenomenon for qubit channels. We further prove that square of the variance of a suitable hermitian operator can act as a non-linear witness of non-Markovianity. We finally show that non-Markovianity is necessary in order to decrease the uncertainty of the states undergoing unital dynamics for qubits.
This provides another method of certifying non-Markovianity.
\end{abstract}

\maketitle

\section{Introduction}
The theory of open quantum systems provides adequate descriptions of general quantum evolution, enabling physical explanations for dissipation, decoherence and other irreversible phenomena due to interaction with noisy environment \citep{alicki,lindblad,gorini,breuer,rivas1,breuerN,alonso}. Usually, the general quantum dynamics can be divided into two specific classes, namely, Markovian and non-Markovian. Markovian evolutions can be broadly identified by those special type of evolutions, for which the information flow from the system to the environment is a one-way traffic. These types of evolutions can be described by divisible quantum maps \cite{RHPreview,BLPreview,Vegareview}. However,  when the system-environment coupling is not sufficiently weak or the environment is non-stationary, we encounter situations where the Born-Markov approximation can not be satisfied.  Therefore, for such evolutions, there exists many such realistic scenarios for which  indivisible dynamics ca
 n also occur,  providing signatures of non-Markovian information back flow\citep{RHP,blp1,bellomo,arend,ban1,ban2}. 

In recent years it has been realized that non-Markovianity (NM) acts as a resource in various quantum mechanical tasks. For example, NM allows perfect teleportation with mixed states \citep{task1}, efficient entanglement distribution \citep{task3}, improvement of capacity for long quantum channels \citep{task2}, and efficient work extraction from an Otto cycle \citep{task4}. For these cases, the accomplishment of the concerned tasks are done by harnessing information backflow, which can be understood as resource inter-conversion. NM can thus be inter-converted via information backflow, into other resources such as entanglement, coherent information and extractable work. It can also be exploited for efficient quantum control \citep{task5}.  Experimentally, non-Markovian system-environment correlations have been demonstrated in trapped-ion \citep{experiment11,experiment8}, nuclear magnetic resonance \citep{experiment12} and photonic \citep{experiment1,experiment2,experiment3,experiment4,experiment10,experiment6,experiment7,experiment9} systems.

In view of the emerging significance of NM in quantum information processing, it has become important to specify whether a given system-environment dynamics possesses non-Markovian traits. In other words, in order to use NM as a resource, one first needs to detect signatures of NM. Thus, formulation
of detection schemes of NM is highly relevant in a  spirit similar to the construction of witness theories of other resources such as quantum entanglement and coherence. Note that  experimentally witnessing signatures of NM is a challenging task, as proposed measures of NM based on information backflow \citep{blp1} are difficult to realize directly in practice. 

Though there  are different approaches in defining non-Markovianity in the quantum domain \citep{RHP,blp1,bellomo,arend} including a recent attempt to construct a generalized description of NM by process tensor formalism \cite{Modi}, in this work we propose an avenue for detecting the signatures of NM, based on the definition of indivisibility of the dynamical map by interlinking it with a fundamental feature of quantum mechanics, {\it viz.}, uncertainty relations. This specific approach allows us to exploit the non-complete positivity spawning from the indivisibility of the channel, which leads to the breaking of uncertainty relations providing a scope to  detect such operations. Our motivation thus stems from the feasibility of experimental realization of uncertainty relations between Hermitian operators, by restricting ourselves within only indivisible quantum operations.

Quantum theory restricts the accuracy of simultaneous measurements, which is well explained by the famous Heisenberg uncertainty relation \cite{Heisenberg}. A stronger lower bound of uncertainty was obtained by Robertson-Schr\"{o}dinger by including an anti-commutator for more generalized pairs of observables\cite{Robertson}. However, for quantum states that are eigenstates of one of the observables, a non-trivial lower bound reflecting incompatibility of the observables may still be obtained by using uncertainty relations
based on sums of variances \cite{Pati,Mac,pati3}. Other forms of uncertainty relations such as entropic uncertainty relation \cite{entropy,wehner}, and fine-grained uncertainty relation \cite{Oppenheim} have also been derived. Experimental investigations of different uncertainty relations have been performed \cite{uncertainty1,uncertainty2,uncertainty3,uncertainty4,uncertainty5,
 uncertainty6,uncertainty8,uncertainty9,uncertainty10,uncertainty11,
 uncertainty12,uncertainty13}
 
 Uncertainty relations have wide range of applicability. They can be used to justify the complex structure of the Hilbert space \cite{Busch,Lathi}, for detecting purity\cite{mal}, entanglement \cite{Guhne,Hofmann,Nha,entropy_entanglement,symplectic,seperability_uncertainty,Nha2,agarwal,landao}, demonstration of EPR-steering \cite{Reid,TP,Ananda,Chowdhury}, security analysis of quantum key distribution \cite{Peres}, etc. Drawing inspiration from the literature on the usage of uncertainty relations to witness quantum correlations, we investigate the possibility of detecting non-Markovianity through uncertainty relations.In a realistic scenario  the overall dynamics is always completely-positive (CP) and hence, uncertainty relations will always hold for the overall dynamics. However, complete-positivity may break down within the intermediate time step for NM evolution. This leads to the violation of uncertainty relations within those intermediate time steps and 
 hence, can be used to detect NM. In the present work we propose the utility of the Robertson-Schr\"{o}dinger (RS) uncertainty relation to detect NM. 

The plan of this paper is as follows. In the next section we construct our theory of non-Markovianity detection via the RS uncertainty violation proposing a linear witness of NM, and verify our theory with some examples. We further present a nonlinear extension of the NM witness and show that it could
lead to the detection of NM dynamics beyond that possible by the linear witness. Then in Section III we discuss a special case of unital dynamics for qubits and propose another  method of detecting non-Markovianity through the dynamics of Robertson-Schr\"{o}dinger uncertainty.  We conclude with a summary of our results and possible implications of our approach in Section IV.

\section{Uncertainty relations for non-Markovianity detection}

We begin with a brief overview of the preliminary background of indivisible  non-Markovian operations and their characterization. 

\subsection{Geometric representation of quantum channels  and CP-divisibity}
 Consider an arbitrary quantum channel, i.e, a completely positive trace preserving map, 
$\Lambda_{\mathcal{N}} (t,t_0) : \rho(t_0) \rightarrow \rho(t)$. Let us introduce a set $\mathcal{D}$ which contains all such quantum channels. Via the channel-state duality \citep{choi,jamil}, we can always find a one to one connection of $\mathcal{D}$ with the set of all Choi-states $\mathcal{F}$, where the Choi-state for any quantum evolution $\Lambda_{\mathcal{N}} (t,t_0)$  is defined by 
\[ \mathcal{C}_{\mathcal{N}}(t,t_0) = \mathbb{I}\otimes\Lambda_{\mathcal{N}}(t,t_0)(|\phi\ket\bra\phi |). \]
Here $|\phi \ket$ is a maximally entangled state of dimension $d \times d$ for a $d$ dimensional system. According to Choi-Jamiolkowski isomorphism \citep{CJ} for complete-positivity (CP) of the dynamics $\Lambda_{\mathcal{N}}(t,t_0)$, it is sufficient to show that the corresponding Choi state $\mathcal{C}_{\mathcal{M}}(t,t_0)\geq 0$, $\forall (t, t_0)$. (Throughout the paper, for Choi states we use $\mathcal{C}_{\mathcal{N}}$ and $\mathcal{C}_{\mathcal{M}}$ interchangeably to mean the same, unless specified explicitly.)


 The channel or dynamical map is called CP-divisible iff it can always be written in the form $\Lambda_{\mathcal{N}} (t,t_0) = \Lambda_{\mathcal{N}} (t, t_1) \circ \Lambda_{\mathcal{N}} (t_1, t_0)$ where $t > t_1 > t_0 ~\forall ~ t, t_1, t_0$. This condition is equivalent to the condition that the dynamics is CP for all time intervals \citep{RHP}. Following Rivas et.al. \citep{RHP}, we shall use the term `Markovian Choi-state' if the corresponding dynamics is CP-divisible. Rest of the Choi states, corresponding to indivisible channels are considered as non-Markovian Choi states.


 A dynamics of Lindblad form is, $\dot{\rho}(t) =\mathcal{L}_t(\rho (t))$ where $\mathcal{L}_{t}(.)= \sum_{i}\Gamma_{i}(t)(L_{i}(.)L_{i}^{\dagger}-\frac{1}{2}\lbrace L_{i}^{\dagger}L_{i},(.)\rbrace).$ Here $\Gamma_{i}(t)$'s are Lindblad coefficients and  $L_{i}$'s are Lindblad operators. A necessary and sufficient condition for CP-divisibility of the dynamics is that $\Gamma_{i}(t)\geq 0 ~\forall (i,t),$ \cite{gorini}.

However, in many practical situations CP-divisibility may break down, leading to the observation of non-Markovian backflow of information. This approach of non-Markovianity via CP-indivisibility has been used extensively in the recent literature \cite{RHP,blp1,bellomo,arend,ban1,ban2}. It is evident from these studies that CP-indivisibility of arbitrary dynamical evolutions leads to the violation of positive semi-definiteness of the corresponding Choi states \citep{choi,jamil}. This leads us to the idea of detecting non-Markovianity by the violation of uncertainty relations.

\subsection{Violation of Uncertainty relations}
  Robertson \cite{Robertson} showed that uncertainty relations can be found using the property of positive semi-definiteness of a quantum state. More recently it has been proven \cite{Nha} that positive semi-definiteness of a state is necessary for uncertainty relations to hold. For example, in a two-dimensional Hilbert space a state is described by a $2\times 2$ matrix $\rho_1 =\left(\begin{array}{cc} a  & c \\ c* & b \end{array}\right) $ such that $Tr[\rho_1]=1$ and $Det[\rho_1] = ab-\vert c\vert^{2} \geq 0$. 
  
  The Robertson-Schrodinger (RS) uncertainty relation for two observables $A$ and $B$ (in arbitrary dimensions) is given by, 
\begin{equation}\label{1}
\Delta^{2} A\Delta^{2} B- \frac{1}{4}\vert \langle[A,B]\rangle\vert^{2} 
-\frac{1}{4}\vert (\langle \lbrace A,B \rbrace\rangle - 2 \langle A \rangle\langle B\rangle)\vert^{2}\geq 0.
\end{equation}
Here the variance of $A$ (similarly for $B$) is defined as $\Delta^{2} A \equiv \langle A^2 \rangle - \langle A\rangle^2$, and all expectation values are calculated over some quantum state $\rho$. The RS relation  holds if $\rho$ is non-negative.

 For a CP divisible evolution $\mathcal{M}$, the corresponding Choi state $\mathcal{C}_{\mathcal{M}}(t+\epsilon,t)\geq 0$, $\forall (t,\epsilon)$, the uncertainty relations will always hold.
However, for some arbitrary CP-indivisible evolution $\mathcal{N}$, $\mathcal{C}_{\mathcal{N}}(t+\epsilon,t)$ can acquire negative eigenvalues within some time interval $\epsilon$ and may lead to the violation of uncertainty relations.
We use this fact to state and prove the following theorem. 

\textbf{Theorem 1 :} The statement of this theorem is twofold.\\ 1) Violation of the Robertson-Schr\"{o}dinger uncertainty relation for the Choi state of a dynamical evolution is sufficient to detect CP-indivisibility. \\2) There always exists a set of hermitian operators, for which RS uncertainty relation will be violated in case of CP-indivisible operations.

\proof Let us consider the derivation \cite{Ballentine} of RS-uncertainty relation given by \eqref{1}. In order to do that, we evaluate $Tr[\mathcal{C}_{\mathcal{N}} TT^{\dagger}]$ for any operator $T= A_0 +(\gamma + i \epsilon) B_0 $, where $A_0 = A - \langle A\rangle$ and $B_0 = B - \langle B\rangle$. $\mathcal{C}_{\mathcal{N}}$ is an arbitrary Choi state over which the averages are calculated. It is easy to show that $TT^{\dagger}$ is always positive semi-definite for any operator $T$. After a few steps of simplification, we get

\begin{eqnarray*}
 Tr[\mathcal{C}_{\mathcal{N}} TT^{\dagger}] 
=&& Tr[\mathcal{C}_{\mathcal{N}} A_0^2] + (\gamma^2 + \epsilon^2)Tr[\mathcal{C}_{\mathcal{N}} B_0^2] \\
&+& \gamma Tr[\mathcal{C}_{\mathcal{N}} \lbrace A_0, B_0 \rbrace] -i \epsilon Tr[\mathcal{C}_{\mathcal{N}}[A_0,B_0]].
\end{eqnarray*}
One can find that the value of $Tr[\mathcal{C}_{\mathcal{N}} TT^{\dagger}]$ is minimum for $\gamma = - \frac{Tr[\mathcal{C}_{\mathcal{N}}\lbrace A_0, B_0\rbrace]}{2Tr[\mathcal{C}_{\mathcal{N}} B_0^2]}$ and $\epsilon =  \frac{i Tr[\mathcal{C}_{\mathcal{N}}[ A_0, B_0]]}{2Tr[\mathcal{C}_{\mathcal{N}} B_0^2]}$. Substituting these, one has
\begin{eqnarray*}
\begin{array}{ll}
\text{min}_{\gamma,\epsilon} 
Tr[\mathcal{C}_{\mathcal{N}} TT^{\dagger}] = \frac{1}{Tr[\mathcal{C}_{\mathcal{N}} B^2] - (Tr[\mathcal{C}_{\mathcal{N}} B])^2 } [(Tr[\mathcal{C}_{\mathcal{N}} A^2] \\- (Tr[\mathcal{C}_{\mathcal{N}} A])^2)(Tr[\mathcal{C}_{\mathcal{N}} B^2] - (Tr[\mathcal{C}_{\mathcal{N}} B])^2)  +\frac{1}{4}\left(  Tr[\mathcal{C}_{\mathcal{N}} [A,B]]\right) ^{2}\\
 - \frac{1}{4}\left(  (Tr[\mathcal{C}_{\mathcal{N}} \lbrace A,B \rbrace ] - 2 Tr[\mathcal{C}_{\mathcal{N}} A ]Tr[\mathcal{C}_{\mathcal{N}} B ])\right) ^{2} ]
\end{array}
\end{eqnarray*}
which represents the left hand side of \eqref{1}.

Now since $TT^{\dagger}$ is always positive; the identity,
\begin{equation}
Tr[\mathcal{C}_{\mathcal{N}} TT^{\dagger}] \geq 0
\end{equation}
always holds, only if $\mathcal{C}_{\mathcal{N}}$ is positive semi-definite.
Using the above identity one can have \eqref{1}. Since $TT^{\dagger}$ is always positive, the identity can break only if $\mathcal{C}_{\mathcal{N}}$ is negative. In other words, the violation of the uncertainty relation is sufficient to know that $\mathcal{C}_{\mathcal{N}}$ is not positive semi-definite, and hence, the dynamics is indivisible.\\

However, in order to show that  violation of the uncertainty relation is not necessary to detect CP-indivisibility, consider that $\mathcal{C}_{\mathcal{N}}$ has a spectral decomposition of the form, 
\begin{equation}\label{spectral}
\mathcal{C}_{\mathcal{N}}= \sum_i \lambda_i^+ |\lambda_i^+\ket \bra \lambda_i^+| +\sum_j \lambda_j^- |\lambda_j^-\ket \bra \lambda_j^-|,
\end{equation} 
 where $\lambda_i^+ \mbox{and} \lambda_j^-$ are the positive and negative eigenvalues respectively. In this case there exists an operator $TT^{\dagger}$ such that $TT^{\dagger}$ is orthogonal to the eigen-basis corresponding to negative eigenvalues of $\mathcal{C}_{\mathcal{N}}$ (for example $TT^{\dagger} = |\lambda_i^+\ket \bra \lambda_i^+|$). In this situation even if $\mathcal{C}_{\mathcal{N}}$ is negative, the uncertainty relation may still hold. 

Now, to prove the second part of the theorem that there always exists  a pair of hermitian operators which will violate the RS uncertainty relation for CP-indivisible Choi states, let us consider the Choi-state in it's spectral decomposition form [\ref{spectral}]. The trace preservation condition yields $\sum_i\lambda_i^++\sum_j\lambda_j^- =1$. This shows us that for any arbitrary choice for the non-Markovian Choi state, there exists at least one positive eigenvalue, to preserve this trace relation. Let us consider one of the Hermitian operators to be $H_1=|\lambda_j^-\ket\bra\lambda_j^-|$, which will give us the negative eigenvalue $\lambda_j^-$, if acted upon the Choi state $\mathcal{C}_{\mathcal{N}}$. For this, the variance will give us $\Delta^2H_1=\lambda_j^--(\lambda_j^-)^2 < 0$. Let us now consider another Hermitian operator $H_2=\sum_{k \neq l}|\lambda_k\ket\bra\lambda_l|$, where $|\lambda_k\ket$ and $|\lambda_l\ket$ can take any values from the spectral decomposition of $\mathcal{C}_{\mathcal{N}}$ with corresponding eigen values $\lambda_k$ and $\lambda_l$ . If there are $n$ number of terms in the spectral decomposition of $\mathcal{C}_{\mathcal{N}}$ then $\langle H_2^2 \rangle = (n-1) \sum_k \lambda_k = n-1$, $\langle H_2 \rangle = 0$ and hence $\Delta ^2 H_2 > 0$. Therefore, the term $\Delta^2H_1 \Delta^2H_2 < 0$, making the left hand side of RS relation \eqref{1} to be  negative, for the pair of operators $H_1$ and $H_2$. For any given Choi state, which is CP-indivisible, i.e., containing at least one negative eigenvalue, we can always find such a pair of $H_1$ and $H_2$. This proves the theorem. \qed


  Recently, it has been shown that one can have convex compact structure of the set of Markovian-Choi states if we restrict ourselves only to those dynamical maps which have Lindblad type generators and in short time interval $(t, t+ \epsilon)$ approximation ($\epsilon$ is sufficiently small) \cite{samya}. In the following proposition, using convex structure of the set of Markovian-Choi states, we discuss another uncertainty like relation, whose violation will sufficiently detect non-Markovian dynamics.
 
\textbf{Proposition 1 :} For some observable $A_{i}$ and some state $\mathcal{C}_{\mathcal{M}}$ written in a  convex combination [i.e. $\mathcal{C}_{\mathcal{M}}= \sum_{k}p_{k}\mathcal{C}_{\mathcal{M}}^{k}$ where $p_{k}\geq 0,\sum_{k}p_{k}=1$] of some Markovian Choi-states, $\mathcal{C}_{\mathcal{M}}^{k}$ within the set of all Markovian Choi-states having Lindblad type generators $(\mathcal{F}^\epsilon_{\mathcal{M}})$,
\begin{equation}
\sum_{i} \Delta^{2}(A_{i})_{\mathcal{C}_{\mathcal{M}}}\geq \sum_{k} p_{k}\sum_{i}\Delta^{2}(A_{i})_{\mathcal{C}_{\mathcal{M}}^{k}},
\end{equation}
holds, where $\Delta^{2}(A_{i})_{X}$ is the variance of $A_i$ over the state $X$ as defined earlier. We call a state "violating proposition 1" if and only if there are no states $\mathcal{C}_{\mathcal{M}}^{k}\in \mathcal{F}^\epsilon_{\mathcal{M}}$ and no $p_{k}$ such that above equation is fulfilled.

\proof This fact is known from a different perspective of entanglement theory \cite{Guhne,Hofmann}. For each $A_{i}$,
\begin{eqnarray}
&&\Delta^{2}(A_{i})_{\mathcal{C}_{\mathcal{M}}} = \sum_{k} p_{k}\langle (A_{i}-\langle A_{i}\rangle_{\mathcal{C}_{\mathcal{M}}})^{2}\rangle_{\mathcal{C}_{\mathcal{M}}^{k}} \nonumber \\
&=& \sum_{k}p_k(\langle A_{i}^2\rangle _{\mathcal{C}_{\mathcal{M}}^{k}} - \langle A_{i}\rangle _{\mathcal{C}_{\mathcal{M}}^{k}}^2 + \langle A_{i}\rangle _{\mathcal{C}_{\mathcal{M}}^{k}}^2 - 2 \langle A_{i}\rangle_{\mathcal{C}_{\mathcal{M}}^{k}}\langle A_{i}\rangle_{\mathcal{C}_{\mathcal{M}}}+\langle A_{i}\rangle_{\mathcal{C}_{\mathcal{M}}}^{2})  \nonumber \\
&=& \sum_{k}p_k(\Delta^{2}(A_{i})_{\mathcal{C}_{\mathcal{M}}^{k}}+(\langle A_{i}\rangle_{\mathcal{C}_{\mathcal{M}}^{k}}-\langle A_{i}\rangle_{\mathcal{C}_{\mathcal{M}}})^{2})  \nonumber \\
& \geq & \sum_{k} p_k\Delta^{2}(A_{i})_{\mathcal{C}_{\mathcal{M}}^{k}}. \nonumber
\end{eqnarray}
Now summing over $i$, we get
\begin{equation*}
\sum_{i} \Delta^{2}(A_{i})_{\mathcal{C}_{\mathcal{M}}}\geq \sum_{k} p_{k}\sum_{i}\Delta^{2}(A_{i})_{\mathcal{C}_{\mathcal{M}}^{k}},
\end{equation*} \qed

 Since the proposition is based on convex structure of the set of Markovian Choi-states $\mathcal{C}_{\mathcal{M}}$, violation of this proposition implies the state $\mathcal{C}_{\mathcal{M}}$ lies outside the convex set and hence must be non-Markovian. In fact this proposition is a sufficient criterion, if we choose $A_{i}$ appropriately.

One may be interested to obtain the lower bound of the above derived uncertainty inequality. From the above equation it is clear that the equality holds for pure Markovian Choi-states, i.e., if the state $\mathcal{C}_{\mathcal{M}}$ does not have any decomposition other than itself. Next, one has to optimize over all such pure Markovian Choi-states. 

In the following proposition, we consider another uncertainty relation, which can be advantageous than the RS uncertainty relation to detect non-Markovianity.

 \textbf{Proposition 2 :} For two hermitian operators $A$ and $B$, 
\begin{eqnarray}\label{new1}
\Delta^{2} A + \Delta^{2} B \geq \vert \langle \left[A, B\right]\rangle\vert,
\end{eqnarray}
where $<(\cdot)> = Tr((\cdot)\mathcal{C}_{\mathcal{M}})$, with $\mathcal{C}_{\mathcal{M}}\in \mathcal{F}_{\mathcal{M}}^{\epsilon}$. 

\proof The proof is straightforward. We know $(\Delta A - \Delta B)^2 \geq 0$ or $\Delta^{2} A + \Delta^{2} B\geq 2 \Delta A\Delta B.$ Now, the Heisenberg uncertainty relation for $A$ and $B$ is given by $ \Delta A\Delta B\geq \frac{1}{2}\vert \langle \left[A, B\right]\rangle\vert $. \qed 

Here we observe that choosing a suitable observable is important, so that the expectation value of the observable corresponding to some non-Markovian Choi-state $\mathcal{C}_{\mathcal{N}}$ becomes negative. Therefore, violation of ``\textbf{Proposition 2}" implies detection of non-Markovianity. The reason to propose this sum uncertainty relation is due to the fact that it can detect non-Markovianity through its violation, in cases where the product uncertainty relation like the RS uncertainty do not show violation. 
Let us elucidate more on this issue of non-Markovianity detection in the context of the following example. 

Consider a non-Markovian dynamics having Choi state $\mathcal{C}_{\mathcal{N}}$, which has at least two negative eigenvalues $\lambda_{1}^{-}$ and  $\lambda_{2}^{-}$, corresponding to the eigen-vectors $\vert\lambda_{1}^{-}\rangle$,$\vert\lambda_{2}^{-}\rangle$ respectively. We construct two projectors $W_{1}=\vert\lambda_{1}^{-}\rangle\langle\lambda_{1}^{-}\vert$ and $W_{2}= \vert\lambda_{2}^{-}\rangle\langle\lambda_{2}^{-}\vert $.The RS uncertainty \eqref{1} for these two observables, gives us
$
\lambda_1^-\lambda_2^- (1- (\lambda_1^- + \lambda_2^-)) \geq 0.
$
Clearly,  equation \eqref{1} is not violated since $\lambda_1^-$ and $\lambda_2^-$ are both negative. However, if one considers the sum uncertainty \eqref{new1}, the LHS  becomes negative, but the RHS being commutator of two orthogonal projectors vanishes. We can see that the violation of equation \eqref{new1} occurs for this case. This example shows that the sum uncertainty relation \eqref{new1} can be advantageous compared to product uncertainty relations for the detection of non-Markovianity.

 \subsection{Examples} 

 In this sub-section, we present various practical examples from quantum channels, to validate the theory we constructed in the above presentation.

 \textit{ Dephasing channel: }
 The Lindblad equation for a pure dephasing channel is given by, $$\frac{d\rho}{dt}=\gamma(t)(\sigma_{z}\rho\sigma_{z}-\rho),$$  
 where $\sigma_z$ is the $z$ component of Pauli matrix and $\gamma(t)$ is defined as,
$$\gamma(t)=\frac{2\lambda\gamma_0 \sinh[tg/2]}{g\cosh[tg/2] + \lambda\sinh[tg/2]}.$$
with $g=\sqrt{\lambda^2-2\gamma_0 \lambda}.$
It has been shown that the dynamics will be non-Markovian when $\gamma(t)$ has a negative region and that happen only when $\gamma_0 > \lambda/2$ \citep{victor}. Here we calculate the LHS of RS-uncertainty relation (Eq. \eqref{1})
$R(t)$ for the Choi-state of this dephasing channel under the small time approximation ($\epsilon\gamma(t) << 1$), and observe that uncertainty relations get violated only  when the Lindblad coefficient $\gamma(t)$ is negative. This confirms that violation of uncertainty relation is  sufficient to show that the dynamics is non-Markovian. Fig. (1) depicts the uncertainty profile as a function of time for the dephasing channel for the observables $S_x=\sigma_{x}\otimes\sigma_{y}$,$S_y=\sigma_{x}\otimes\sigma_{x}$. Note that, in figure we consider dimensionless quantity $\bar{t}=t/\kappa$ and $\bar{\Gamma}(t)=\kappa\gamma(t)$ for time and Lindblad coefficient respectively, where $\kappa$ is a constant of the dimension of time.   
\begin{figure}[htbp]\label{fig1}
  \begin{center}
      \includegraphics[height=6 cm, width=7cm]{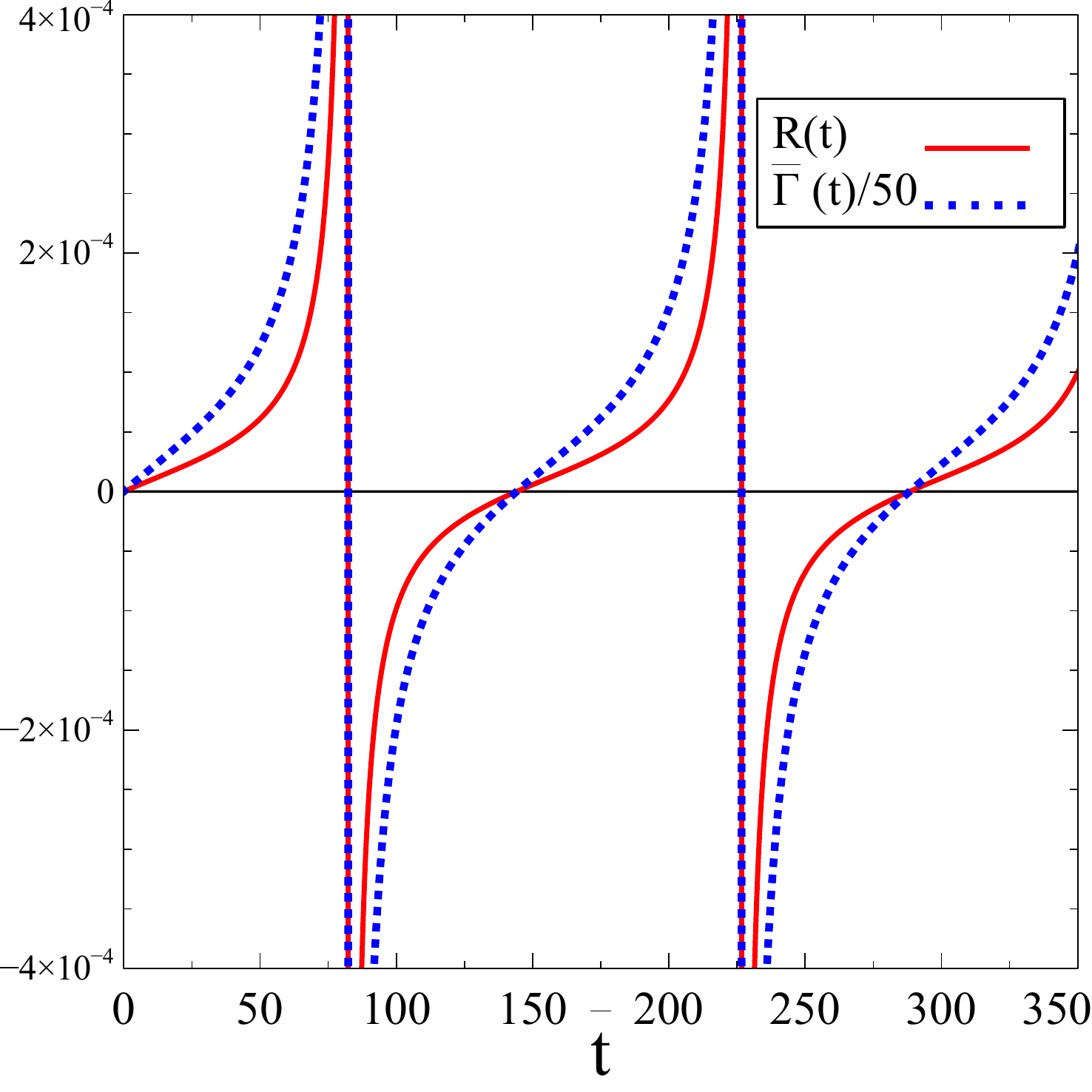}
\caption{{(Colour online) Plot for the RS-uncertainty quantity $R(t)$ for a Choi-state and the Lindblad-coefficient ($\bar{\Gamma}(t)$) with dimensionless time $\bar{t}$ for the Choi-state of the dephasing channel. Clearly, the RS-uncertainty relation is violated (i.e, R(t)$< 0$) only  where $\bar{\Gamma}(t)$ is negative.}}
\end{center}
\end{figure}

 \textit{Spin bath model:}
 Here we  use the generalized non-Markovian evolution for a spin-bath model. This type of system has drawn significant interest in recent studies \citep{spin1,spin2,spin3,samya2}. The model considered here in the present work consists of a central spin interacting uniformly with a collection of mutually non-interacting and completely unpolarized spins, considered as the environment. An exact canonical master equation of the Lindblad form for such a model has been derived earlier \cite{exact}. From the master equation of the Lindblad form it was shown that the reduced dynamics of the central spin exhibits non-Markovian features throughout the evolution. The dynamical equation for the mentioned reduced evolution is given by
\begin{eqnarray}\label{new2}
 \dot{\rho}(t) &= \frac{i}{\hbar} U(t)\left[\rho (t), \sigma_{z}\right] + \Gamma_{deph}(t)\left[\sigma_{z}\rho (t)\sigma_{z}-\rho (t)\right] \nonumber \\ 
 &+ \Gamma_{dis} (t)\left[\sigma_{-}\rho (t)\sigma_{+} -\frac{1}{2}\left\lbrace \sigma_{+}\sigma_{-},\rho (t)\right\rbrace \right] \nonumber \\
 &+ \Gamma_{abs}(t)\left[\sigma_{+}\rho (t)\sigma_{-}-\frac{1}{2}\left\lbrace \sigma_{-}\sigma_{+},\rho (t)\right\rbrace \right] ,
\end{eqnarray}
where $\sigma_{\pm}=\frac{\sigma_{x}\pm i\sigma_{y}}{2}$ and $\Gamma_{dis}(t),\Gamma_{abs}(t), \Gamma_{deph}(t)$ are the rates of dissipation, absorption, dephasing processes respectively, and $U(t)$ corresponds to the unitary evolution. The uncertainty profile for this particular evolution is depicted in Fig. (2), showing the detection of non-Markovianity through violation of uncertainty. Note that, similar to the case of dephasing channel, all the quantities in the plot are made dimensionless, by scaling with the parameter $\kappa$.

\begin{figure}[htbp]\label{fig3}
  \begin{center}
      \includegraphics[height=6 cm, width=7cm]{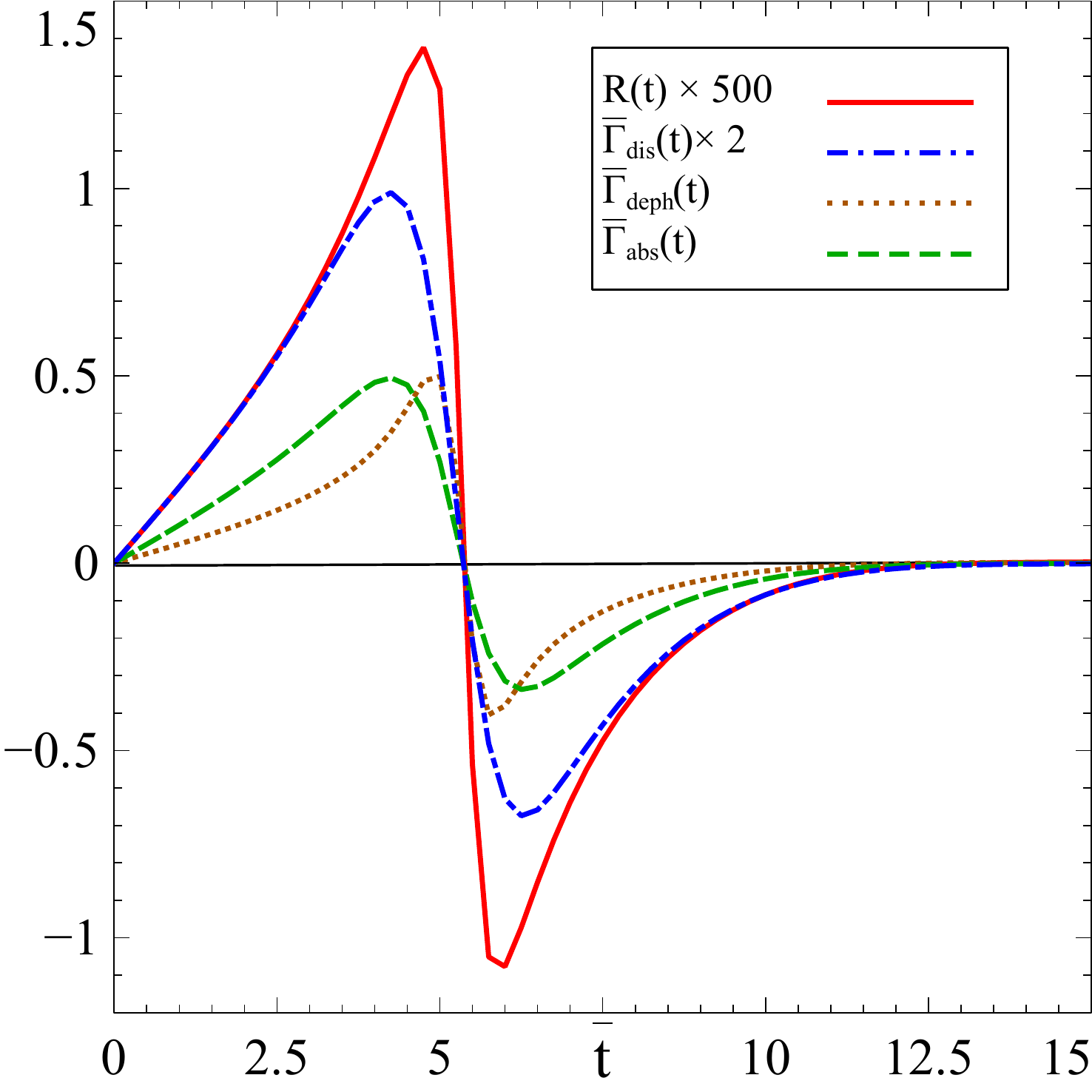}
\caption{{(Colour online) Plot for RS-uncertainty relation (R(t)) for a Choi-state and various dimensionless Lindblad-coefficients ($\bar{\Gamma}_{dis}(t),\bar{\Gamma}_{abs}(t), \bar{\Gamma}_{deph}(t)$) with time $\bar{t}$ for the Choi-state of spin-bath model. Clearly RS-uncertainty relation is violated (i.e, R(t)$< 0$) only at those regions where at least one of the Lindblad-coefficients is negative.}}
\end{center}
\end{figure}

\subsection{Variance as a Non-linear witness of non-Markovianity}
 Recently, it has been shown that for  dynamical maps having Lindblad type generators, one can have a resource theory of non-Markovianity where divisible operations are the free operations and the Choi-states corresponding to the free operations are the free states \cite{samya}. It was  shown that for a sufficiently small $\epsilon$, the set of all Markovian Choi-states  having Lindblad type generators will take a convex and compact form $(\mathcal{F}^\epsilon_{\mathcal{M}})$. In a following work \cite{bihalan},   a theory of linear witnesses was further established for detecting non-Markovianity. It was shown that the set of Choi states for divisible operations does not form a polytope. Consequently, we can surmise that linear witnesses will not be sufficient for optimal detection of non-Markovianity, and hence, non-linear corrections of linear witnesses are in order. The possibility of detecting non-Markovianity via violation of uncertainty relations gives us one such opportunity. As we have mentioned earlier,  one can construct NM witness \cite{bihalan} by utilizing the fact that the  Markovian Choi states having Lindblad generators form a convex and compact set ($\mathcal{F}^\epsilon_{\mathcal{M}}$) under the small time interval approximation \cite{samya}. A hermitian operator $W$ is said to be a non-Markovian witness if the following criteria hold:\\
1.$Tr(W\mathcal{C}_{\mathcal{M}})\geq 0 ~ \forall \mathcal{C}_{\mathcal{M}} \in \mathcal{F}^\epsilon_{\mathcal{M}}$,\\
2. There exists at least one NM Choi-state $\mathcal{C}_{\mathcal{N}}$ such that $Tr(W\mathcal{C}_{\mathcal{N}})< 0$.

 Let $\mathcal{C}_{\mathcal{M}}$ be a Choi-state. Since $\mathcal{C}_{\mathcal{M}}$ is hermitian, one can have it's spectral decomposition of the form
$$\mathcal{C}_{\mathcal{M}}= \sum_{i}\lambda_{i}P_{i},$$
where $P_{i}$s are orthogonal projectors corresponding to the eigenvalues $\lambda_{i}$. If the state is Markovian (CP-divisible) then $\mathcal{C}_{\mathcal{M}}$ is a valid state as it has all non-negative eigenvalues. Hence, $Tr(\mathcal{C}_{\mathcal{M}}P_{i})\geq 0$, $\forall i$. However, if the operation is non-Markovian, then  $Tr(\mathcal{C}_{\mathcal{N}}P_{i})< 0$ for at least one $i$. So, orthogonal projectors serve as witness.

 \textbf{Proposition 3 :} Corresponding to a projective NM witness $W_{i}$, it's variance $\Delta^{2}W_{i}$ is also a witness.
\proof The variance of $W_{i}$ over some state $\mathcal{C}_{\mathcal{N}}$ as defined earlier is 
$\Delta^2 W_i = Tr\left[\mathcal{C}_{\mathcal{N}}. W_{i}^{2}\right] - (Tr\left[\mathcal{C}_{\mathcal{N}}. W_{i}\right])^{2}$.
Since $W_{i}$ is a projector corresponding to the eigenvalue say $\lambda_{i}$, $W_{i}= \vert \lambda_{i}\rangle\langle \lambda_{i}\vert$, so $W_{i}^{2}= W_{i}= \vert \lambda_{i}\rangle\langle \lambda_{i}\vert $. If $\mathcal{C}_{\mathcal{N}}$ is Markovian, then it has all non negative eigenvalues. We know that variance for such a state cannot be negative, i.e.,
$$\Delta^{2}(W_{i}) \geq 0, \forall \mathcal{C}_{\mathcal{N}}\in \mathcal{F}^\epsilon_{\mathcal{M}}.$$
But if the operation is non-Markovian,  $Tr\left[\mathcal{C}_{\mathcal{N}}W_{i}\right]< 0$ for at least one $i$, 
say for $i=1$, and  then the variance for the witness (projector corresponding to the negative eigenvalue) becomes
$$\Delta^{2}(W_{1})= \lambda_{1}- (\lambda_{1})^{2},$$ which is always negative. Therefore, $\Delta^2(W_i)$ satisfies the conditions 1 and 2, and hence, serves as a non-linear witness. \qed 

In fact it is clear that $\Delta^2(W_i)$ is an improvement over the actual projective witness $W_i$, since the variance contains an additional negative term. There is another corollary which can be drawn as a consequence of \textbf{Proposition 3}. 

\textit{Corollary:} Variance of some suitable hermitian operator $\mathcal{H}$ over Choi states can be considered as a non-linear witness of non-Markovianity.

\proof The variance of a hermitian operator $\mathcal{H}$ can be expressed as \[\Delta^2 (\mathcal{H}) = Tr[\mathcal{H}^{\dagger}\mathcal{H}\mathcal{C}_{\mathcal{N}}]-(Tr[\mathcal{H}\mathcal{C}_{\mathcal{N}}])^2\]
Now, similar to the proof of \textbf{Proposition 3}, it can be shown that
\[
\begin{array}{ll}
1.~~\Delta^2 (\mathcal{H}) \geq 0~~\mbox{for all Markovian Choi states}~~ \mathcal{C}_{\mathcal{M}},\\
2.~~\mbox{There exists at least  one non-Markovian Choi state}~~ \mathcal{C}_{\mathcal{N}},\\
~~~~~ \mbox{for which}~~ \Delta^2 (\mathcal{H})< 0.
\end{array}
\]
This proves the corollary. \qed

\section{Uncertainty based Non-Markovianity quantifier for unital dynamics of qubits}
In this section we consider the case of qubit channels and unital dynamics which have been studied recently in several experimental works \cite{experiment11,experiment8,experiment12,experiment1,experiment2,experiment3,experiment4,experiment10,experiment6,experiment7,experiment9}.   Here the RS-uncertainty  function calculated over a physical state $\rho(t) =  \Lambda (t) \rho(0)$ (note that this is the time evolved state, not the Choi-state). We show  that if we restrict ourselves  to  unital dynamics, then the RS-uncertainty quantity is a monotonically increasing function and can only decrease in the regions where CP divisibility breaks down.

 \noindent \textbf{Proposition 4 :} Non-Markovianity is necessary to decrease RS-uncertainty of time-evolving physical states undergoing unital dynamics.\\
\proof  It has been shown that the RS-uncertainty relation can be used to detect purity and mixedness of a state\cite{mal}.  
The RS-uncertainty relation for a physical state $\rho$ may be written as
\begin{equation}
\Re (A, B, \rho)\geq 0,
\end{equation}
where $\Re (A, B, \rho)$ is defined as,
\begin{equation}
\Re (A, B, \rho) \equiv \Delta^{2} A \Delta^{2} B - \vert \frac{\langle\left[A,B\right]\rangle}{2}\vert^{2} - \vert (\frac{\langle \lbrace A,B \rbrace\rangle}{2}-\langle A\rangle\langle B\rangle)\vert^{2}.
\end{equation}
On the other hand, linear entropy is defined as 
\begin{equation}
\label{9}
S_{l}(\rho)=\frac{d}{d-1}\left[1- Tr\left[\rho^{2}\right]\right],
\end{equation}
where, a qubit $\rho$ may in general be expressed as $\rho=\frac{\bf{I} +\widehat{n}.\widehat{\sigma}}{2}.$

By choosing $A= \widehat{r}.\widehat{\sigma}$ and $B= \widehat{t}.\widehat{\sigma}$,
it has been  shown \citep{mal} that 
\begin{equation}
\Re (A, B, \rho)=\left[1-\left(\sum_{i} r_{i}t_{i}\right)^{2}\right] S_{l}(\rho).
\end{equation}
 It follows from Eq[\ref{9}] that,
\begin{equation}
\frac{d}{dt}S_{l}(\rho)= - 2 \frac{d}{d-1}Tr\left[\rho \dot{\rho}\right].
\end{equation}
 Further, considering unital dynamics  of Lindblad operators \citep{samya,exact,lider}, it can be shown that,
\begin{equation}
\frac{d}{dt}S_{l}(\rho)= \frac{d}{d-1} \sum_i \Gamma_i(t)Q_{i}(t),
\end{equation}
where $Q_{i}(t) =  \parallel [V_{i},\rho (t)]\parallel_{HS}^{2}$ with $V_i$ the Lindblad operator for an arbitrary unital channel and $\parallel X\parallel_{HS}=\sqrt{Tr[X^{\dagger}X]}$ is the Hilbert-Schmidt norm. $Q_{i}(t)$ is a known measure of quantumness \citep{vedral,quan1}. Combining the above results, here we get
\begin{eqnarray}
\frac{d\Re}{dt}
& =& \frac{d}{d-1}\left[1-\left(\sum_{i} r_{i}t_{i}\right)^{2}\right] \sum_{i}\Gamma_{i}(t)Q_{i}(t),
\label{unital}
\end{eqnarray}
where, $\sum_i r_i t_i < 1$.
 Since $Q_{i}(t)$ is always positive, uncertainty will always increase  unless $\Gamma_{i}(t)$ becomes negative. Hence, non-Markovianity is necessary to decrease uncertainty for any unital dynamics of a qubit. \qed

 In order to avoid any dependency of the above expression over observables $A$ and $B$, one may choose a priori  orthogonal operators i.e, $\widehat{r}.\widehat{t} = 0$.
We plot the evolution of RS-uncertainty quantity with time in Fig. (3), for a initial qubit $\rho(0)=\frac{1}{2}\vert + \rangle \langle +\vert $ (where $\vert +\rangle=\frac{\vert 0\rangle +\vert 1\rangle}{\sqrt{2}}$), evolving under unital dynamics (when $\Gamma_{abs} = \Gamma_{dis}$) for the spin-bath model given by equation \eqref{new2} with observables $A = \sigma_x$ and $B = \sigma_y$. It is clear from the plot that the RS-uncertainty  decreases only when the CP-divisibility breaks down. 

\begin{figure}[htbp]\label{fig4}
  \begin{center}
      \includegraphics[height=6 cm, width=7cm]{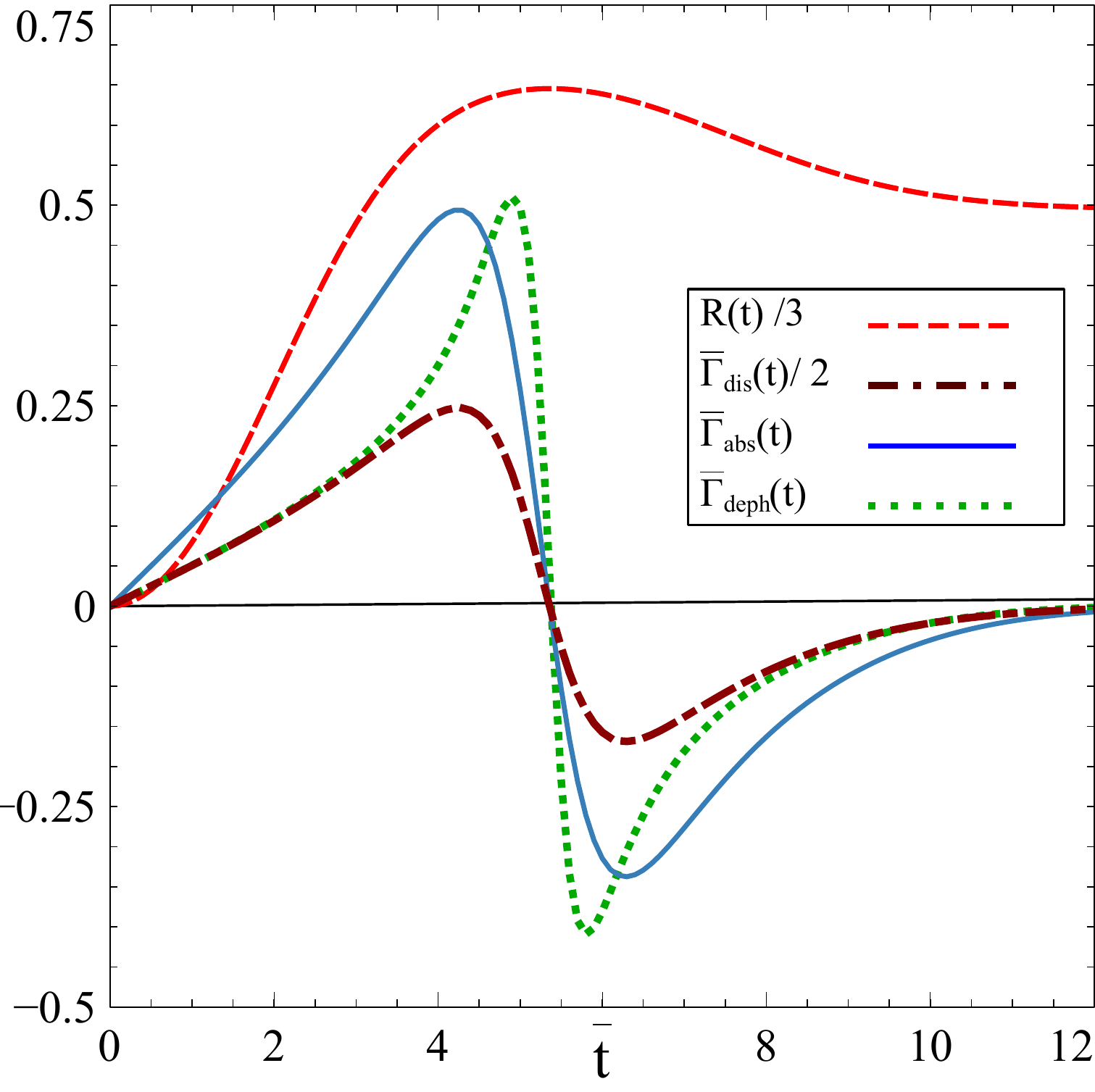}
\caption{{(Colour online) Plot for RS-uncertainty relation for a time-evolving physical state and various Lindblad-coefficients ($\bar{\Gamma}_{dis}(t),\bar{\Gamma}_{abs}(t), \bar{\Gamma}_{deph}(t)$) with time $\bar{t}$ for a qubit undergoing unital dynamics given by the spin bath model. It is seen  that negativity of Lindblad-coefficient (and hence non-Markovianity) is necessary  in order to decrease uncertainty for this unital dynamics. All quantities are dimensionless.}}
\end{center}
\end{figure}

The above findings show that the decrease of $\Re$ for time-evolving physical states signifies information backflow. Based on this, we propose a non-Markovian quantifier for unital qubit dynamics, which can be defined as,
\beq\label{newA}
\mathcal{N}= -\int_{\frac{d \Re }{dt}< 0} \frac{d\Re}{dt}dt.
\eeq
This measure amounts to a certain quantification of total amount of non-Markovianity.  As a further remark, choosing $\widehat{r}$ and $\widehat{t}$ as orthogonal vectors, our proposed measure of non-Markovianity turns out to be identical to the  measure based on purity \cite{exact}.

\section{Conclusions}
To summarize, in this work our goal is to develop frameworks for  detection of non-Markovianity using the mechanism of quantum uncertainty relations. In particular, we formulate two different ways to detect signatures of non-Markovianity in open system dynamics. First, we establish that violation of uncertainty relations for Choi states corresponding to the evolution dynamics is sufficient to detect non-Markovianity. Our formalism is exemplified in the context of  dephasing and spin-bath models  for which non-Markovian evolution can be certified by the violation of the Robertson-Schrodinger uncertainty relation. Based on our approach, we propose certain linear and non-linear witnesses of non-Markovianity. We further present an example to clarify the advantage of the sum uncertainty relation for non-Markovianity detection. Secondly, adopting another approach of using uncertainty relations, we show that for unital dynamics non-Markovianity is necessary to decrease the RS-uncertainty for the time-evolving physical states. This enables us to propose a  new non-Markovianity quantifier for unital dynamics of qubits. 

 Before concluding, it may be worthwhile to discuss the feasibility of experimental verification of the approaches discussed here. Variances of observables have been measured using optical set-ups in several experimental schemes  \cite{uncertainty6,uncertainty11, uncertainty13}. On the
other hand, there has been considerable recent activity towards experimental realization of non-Markovianity under dephasing optical channels \cite{experiment6,experiment7,experiment9,experiment10,experiment11}. In light of these results, it may be feasible to measure the RS uncertainty corresponding to the physical states undergoing unital dynamics, as discussed in Section III, in order to conclude whether the dynamics is non-Markovian. It is also important to mention that in order to compute the uncertainty quantity, we do not need to do a full process tomography as required for several other methods of detecting non-Markovianity \cite{QFI,RHP,blp1,bellomo,arend}. Therefore, in case of determining the nature of the specific quantum evolution experimentally, our method could be  more efficient and cost effective.  It is thus evident that with further development our framework proposed in Section III  should provide an avenue for experimental investigation of non-Markovian evolution in a 
quantitative manner.

\begin{acknowledgments}
 AGM and SB thanks Bihalan Bhattacharya of S.N. Bose Centre for Basic Sciences, Kolkata for useful discussions. SB thanks SERB, DST, Government of India for financial support. ASM acknowledges support from the DST project DST/ICPS/QuEST/2019/Q98.

\end{acknowledgments}

\bibliographystyle{apsrev4-1}
\bibliography{nonmarkovianity_uncertainty}
\end{document}